# A Local Tree Structure is Not Sufficient for the Local Optimality of Message-Passing Decoding in Low Density Parity Check Codes

Weiyu Xu
Department of Electrical Engineering
California Institute of Technology,MC 136-93
Pasadena, CA 91125 USA
Email: weiyu@caltech.edu

*Abstract*—We address the problem,'Is a local tree structure sufficient for the local optimality of message passing algorithm in low density parity check codes?'.It is shown that the answer is negative. Using this observation, we pinpoint a flaw in the proof of Theorem 1 in the paper 'The Capacity of Low-Density Parity-Check Codes Under Message-Passing Decoding' by Thomas J. Richardson and Rüdiger L. Urbanke[1]. We further provide a new proof of that theorem based on a different argument.

## I. INTRODUCTION

Message passing algorithm is an efficient and powerful algorithm in decoding low density parity check codes. In the paper 'The Capacity of Low-Density Parity-Check Codes Under Message-Passing Decoding' by Thomas J. Richardson and Rüdiger L. Urbanke, the authors present an extraordinary way of analyzing the performance of low density parity check codes by proving the concentration results of the local cycle-free structure (namely tree-like structure) and related density evolution. We will use the same notations and definitions as in [1]. In the proof of Theorem 1 (Monotonicity for Physically Degraded Channels), the authors use the following observation:

"... [assuming tree-like neighborhoods $\mathcal{N}_{\vec{e}}^{2\ell}$] for a belief-propagation decoder, the sign of the message sent along edge $\vec{e}$ in the $\ell th$ iteration is equal to the the estimate of an maximum-likelihood estimator based on the observations in $\mathcal{N}_{\vec{e}}^{2\ell}$[2]".

This observation is a key argument in establishing Theorem 1, which was used to validate the error performance analysis and threshold determination in the following part of [1].Beside this,the authors also use this observation for the explanation and analysis of belief-propagation decoder in [1]. However, being kind of surprising, this observation is not true in general. We will give a closer look at this problem and discuss its consequences.

## II. MESSAGE PASSING DECODING FOR A LOCAL TREE STRUCTURE: A CLOSER LOOK

It is well known that when message passing algorithm is applied to a tree-structured tanner graph, the resulting estimate for a variable node in the tree structure is equivalent to the maximum-likelihood estimation of that variable node based on the observations of the variable nodes involved in that tanner graph. This result was carried on in [1] to a local tree-structured part of a certain tanner graph and it was assumed that the same result held true for a tree-structured local part conditioned on the local observations. As we will see, we do not have this in general. Although the constraints specified by the local tree-structured tanner graph are the constraints the variable nodes locally involved satisfy, those constraints may *not* be the *only* constraints that the involved variable nodes *must* satisfy. In other words, there may be *implicit* additional constraints for that local tree-structured neighborhood if we infer a complete tanner graph for that local part. Those additional constraints may bring cycles to the local tree-structured part and undermine the optimality of belief-propagation decoder conditioned on local observations. Here is the detailed theoretical analysis and examples.

We consider a LDPC code of length $n$ with parity check matrix $\mathbf{H}$, which is of dimension $m \times n$. The codebook for this LDPC is denoted by $C$. Without losing generality, we assume this code to be a regular low density parity check code, in which each variable node has a degree of $d_v$ and each parity check node has a degree of $d_c$. Each codeword in the codebook is transmitted with equal probability through a binary-input memoryless channel and we denote the transmitted codeword as $\mathbf{x}^n$ and the received symbols as $\mathbf{y}^n$. Now let us consider the directed neighborhood of depth 2 of the directed edge $\vec{e} = (v, c)$, as shown in Figure 1, which is the same as Figure 2 in [1].

Suppose that we have a local observation $\mathbf{y}_\mathcal{I}$ of the received vector $\mathbf{y}^n$, where $\mathcal{I}$ is the set of indexes $i$ such that $y_i$ is the observation of the transmitted bit $x_i$ corresponding to a variable node involved in the local tree-like neighborhood $\mathcal{N}_{\vec{e}}^{2\ell}$. Now we are in a position to derive the maximum-likelihood estimation for the variable node $v$ based on the local observation $\mathbf{y}_\mathcal{I}$.

$$P(\mathbf{y}_\mathcal{I}|x_i = 1) = \frac{P(x_i = 1, \mathbf{y}_\mathcal{I})}{P(x_i = 1)} \quad (1)$$

$$P(\mathbf{y}_\mathcal{I}|x_i=0) = \frac{P(x_i=0,\mathbf{y}_\mathcal{I})}{P(x_i=0)} \qquad (2)$$

Obviously, we have the following equations:

$$P(x_i=1,\mathbf{y}_\mathcal{I}) = \sum_{x_i=1,\mathbf{x}^n\in\mathcal{C}} P(\mathbf{x}^n)P(\mathbf{y}_\mathcal{I}|\mathbf{x}^n) \qquad (3)$$

$$P(x_i=0,\mathbf{y}_\mathcal{I}) = \sum_{x_i=0,\mathbf{x}^n\in\mathcal{C}} P(\mathbf{x}^n)P(\mathbf{y}_\mathcal{I}|\mathbf{x}^n) \qquad (4)$$

Since the channel is a binary-input memoryless channel, we have

$$P(\mathbf{y}_\mathcal{I}|\mathbf{x}^n) = P(\mathbf{y}_\mathcal{I}|\mathbf{x}_\mathcal{I}) \qquad (5)$$

In (5), we denote $\mathbf{x}_\mathcal{I}$ as the set of transmitted bits corresponding to the index set $\mathcal{I}$. We further denote $\mathcal{C}_\mathcal{I}$ as the sub-codebook for $\mathbf{x}_\mathcal{I}$. Combining (3),(4) and (5), we have

$$P(\mathbf{y}_\mathcal{I}|x_i=1) \propto \sum_{x_i=1,\mathbf{x}_\mathcal{I}\in\mathcal{C}_\mathcal{I}} P(\mathbf{x}_\mathcal{I})P(\mathbf{y}_\mathcal{I}|\mathbf{x}_\mathcal{I}) \qquad (6)$$

$$P(\mathbf{y}_\mathcal{I}|x_i=0) \propto \sum_{x_i=0,\mathbf{x}_\mathcal{I}\in\mathcal{C}_\mathcal{I}} P(\mathbf{x}_\mathcal{I})P(\mathbf{y}_\mathcal{I}|\mathbf{x}_\mathcal{I}) \qquad (7)$$

The derivation in (6) (7) follows from the fact each codeword in the sub-codebook $\mathcal{C}_\mathcal{I}$ induced from the original code $\mathcal{C}$ has equal probability.

Without loss of generality, we index the variable nodes of the bottom layer in Fig.1 as $1,2,\ldots,10$ from the left side to the right side. Then the index set $\mathcal{I} = \{v\}\bigcup\{1,2,\ldots,10\}$. The belief propagation decoder works perfectly if the codeword space $\mathcal{C}_\mathcal{I}$ is a linear space whose null space is specified by the lower 2 parity check nodes in Fig.1 [3]. However, the linear space of the codebook $\mathcal{C}_\mathcal{I}$ may be a proper subspace of the linear space described in Fig.1.

For example, suppose we have additional constraints

$$x_1 \oplus x_{11} \oplus x_{12} \oplus x_{13} \oplus x_{14} \oplus x_{15} = 0; \qquad (8)$$
$$x_2 \oplus x_{11} \oplus x_{12} \oplus x_{13} \oplus x_{14} \oplus x_{15} = 0. \qquad (9)$$

Here $x_{11},\ldots,x_{15}$ are some variable nodes that do not appear in the local neighborhood $\mathcal{N}_{\vec{e}}^{2\ell}$. Although we do not have explicit local cycles, we have an implicit constraint $x_1 = x_2$ for $\mathcal{N}_{\vec{e}}^{2\ell}$. Since the subcode linear space is a proper space of the linear space specified only by the local tree structure, the prior probability of the transmitted subcodeword is not uniform over the linear space specified by $\mathcal{N}_{\vec{e}}^{2\ell}$. In this case, the belief-propagation decoder will not necessarily give maximum-likelihood estimation of the bit corresponding to the variable node $v$.

## III. COMMENTS ON THE PROOF OF THEOREM 1 IN [1]

Let us now look at Theorem 1 in [1]:

"Let $W$ and $W'$ be two given memoryless channels that fulfill the required channel symmetry conditions. Assume that $W'$ is physically degraded with respect to $W$. For a given code and a belief-propagation decoder, let $p$ be the expected fraction of incorrect messages passed at the $\ell$th decoding iteration assuming tree-like neighborhoods and transmission over channel $W$ and let $p'$ denote the equivalent quantity for transmission over Channel $W'$. Then $p \le p'$."

Here is an outline of the original proof of Theorem 1 in [1]:

$(A)$ "Since the transmitted bit associated with variable node $v_i$ has uniform a priori probability, an ML estimator is equal to a maximum a posterior estimation, which is known to yield the minimum probability of error of all estimators based on the same observation";

$(B)$ The maximum a posterior estimator based on the received observation $R$ by sending a randomly chosen codeword through the channel $W$ is superior to the maximum a posterior estimator based on the received observation $R'$ by sending $R$ through an auxiliary channel $Q$;

$(C)$ "The claim now follows by observing that for a belief-propagation decoder decoder, the sign of the message sent along edge $\vec{e}$ in the $\ell$th decoding iteration is equal to the estimate of an ML estimator based on the observation in $\mathcal{N}_{\vec{e}}^{2\ell}$ [2]"

However, two of these claims are inherently flawed.

$(A)$ With a local tree-like neighborhood, we do *not* necessarily have uniform prior probability for the transmitted bit $v_i$ and the sign of the message sent along edge $\vec{e}$ of belief-propagation decoder is not necessarily equal to maximum a posterior estimator even if we assume that "the sign of the message sent along edge $\vec{e}$ in the $\ell$th decoding iteration is equal to the estimate of an ML estimator based on the observation in $\mathcal{N}_{\vec{e}}^{2\ell}$[2]".

$(C)$ The belief propagation decoder is *not* necessarily a maximum-likelihood estimator based on local observations even for a local tree-like neighborhood following the results of Section II.

We now give an example to explain claim (A) by considering an LDPC code with $d_v=3, d_c=5$ and a tree-like neighborhood of depth 2, which is similar to Fig.1 except that each check node has 4 descendent variable nodes. So in this case, we have $\mathcal{I} = \{v\}\bigcup\{1,2,\ldots,8\}$. Similar to the example given in Section II, we can have the following implicit equations even if we do not have cycles in this neighborhood,

$$x_1 = x_2, x_3 = x_4; \qquad (10)$$
$$x_5 = x_6, x_7 = x_8; \qquad (11)$$

We can easily infer from these additional constraints that the variable node $v$ can only take the value 0 from GF(2)

so that it does not have uniform a prior probability. Thus the argument that maximum-likelihood estimator is equivalent to the maximum a posterior probability is not valid in general. So a maximum-likelihood estimator may not yield the minimum probability of error and thus may lose the monotonicity of probability of error for physically degraded channels.One may argue that this case of a redundant bit can not happen in a good low density parity check code. However, in the analysis of a random low density parity check codes generated from a random graph as in [1], one can not exclude this case to happen.

Now comes the question whether Theorem 1 from [1] is true. It turns out that we can establish this theorem by relating the error probability analysis of belief propagation decoder for a tree neighborhood with a belief-propagation decoder for a global tree-structured tanner graph.

Consider a variable node $v_i$ with a local tree neighborhood $\mathcal{N}_{\vec{e}}^{2\ell}$ and a variable node $v_i'$ with a global tree neighborhood $\mathcal{N}_{\vec{e}}^{2\ell}$, which means the constraints specified by the tree structure $\mathcal{N}_{\vec{e}}^{2\ell}$ are the only constraints.The validity of Theorem 1 can be established through the following arguments.

- $(A'')$ For a fixed transmitted all-zero codeword and a certain channel realization,the sign of the message sent along edge $\vec{e}$ in the $\ell$th decoding iteration is equal to the sign of the maximum-likelihood estimate of the bit $v_i'$ based on the observation in $\mathcal{N}_{\vec{e}}^{2\ell}$.Thus the occurrence of belief-propagation decoding error of $v_i$ and $v_i'$ are the same conditioned on the same transmitted codeword and the same channel realization.
- $(B'')$ Due to the channel symmetry conditions, we conclude that the error probability of belief propagation decoding for $v_i$ is the same as the error probability of belief propagation decoding for $v_i'$, which is equal to that of the maximum-likelihood decoding for $v_i'$ based on the observation within the neighborhood of $\mathcal{N}_{\vec{e}}^{2\ell}$.
- $(C'')$ For $v_i'$, the maximum-likelihood estimator based on the observation in $\mathcal{N}_{\vec{e}}^{2\ell}$ is equivalent to the maximum a posterior estimator, for which the estimation of $v_i'$ based on $R$ has no larger error probability than the estimation of $v_i'$ based on $R'$.Combining $(B'')$ and $(C'')$, we have Theorem 1 in a local tree neighborhood whether we have implicit local cycles or not.

To summarize, the original proof too optimistically evaluated a belief-propagation decoder to be the best estimator of a variable node in a local tree neighborhood, which easily implied the superiority of the belief-propagation algorithm for non-degraded channel.But this argument fails since a belief propagation decoder is not always the best estimator in a local tree neighborhood. However,the new proof adds a new element of establishing the equivalence of the error probability of belief propagation decoding in a local tree neighborhood with that of belief-propagation decoding in a global tree neighborhood through the memoryless channel symmetries.Although the belief propagation decoder may not be the maximum-likelihood estimator nor the maximum-a-posterior estimator, it preserves the monotonicity of physical channels by this equivalence. This outcome is largely due to the simultaneous potential suboptimality of belief-propagation decoders both in nondegraded and in physically degraded channels.

## IV. CONCLUSIONS

In this paper,we point out the subtle fact that a local tree structure is not sufficient for the local optimality of message-passing algorithm for low density parity check codes.Based on this, we discuss some subtle but serious flaws in the original proof of Theorem 1 "Monotonicity for Physically Degraded Channels" in [1]. We further provide a new proof for Theorem 1 based on an equivalence of error performance of message-passing algorithm for a local tree neighborhood with the error performance for a global tree neighborhood.

## ACKNOWLEDGMENT

We thank Professor R.J.McEliece for helpful discussions.